\documentstyle[12pt]{article}
\input psfig

\def\A{{\cal A}}

\title{Computational Geometry Column 32}
\author{Joseph O'Rourke%
\thanks{
Department\ of\ Computer\ Science,
Smith College,
Northampton, MA 01063,
USA.
{\tt orourke@cs.smith.edu}.
Supported by NSF grant CCR-9421670.}
}
\date{}
\begin{document}
\bibliographystyle{alpha}
\maketitle
\pagestyle{empty}
\thispagestyle{empty}

\begin{abstract}
The proof of Dey's new $k$-set bound is illustrated.
\end{abstract}
\begin{figure}[htbp]
\begin{center}
\ \psfig{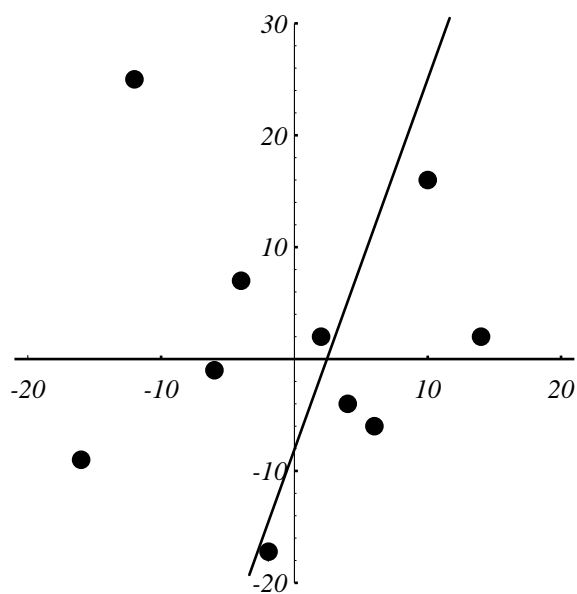}
\end{center}
\caption{The line shown determines two $5$-sets.
There are a total of twenty-four $5$-sets in this set of $n=10$ points.
}
\label{dey.points}
\end{figure}

A {\em $k$-set\/} of a set of $n$ points $P$ in the plane is a set of 
exactly $k$ points of $P$ contained in an open halfplane.
The maximum number of $k$-sets as a function of $n$ and $k>0$
is an important quantity, but a gap in the known bounds between
$\Omega(n \log k)$ and $O(n k^{1/2})$ stood for over $25$ years.
The only progress was a tight bound of $nk$ 
on the number of ${\le}k$-sets,
and a slight lowering of the upper bound for $k$-sets
to $O(n k^{1/2}/\log^* k)$.\footnote{
	The first result, which will play a role below, 
	is due to Alon and Gy\H{o}ri~\cite{ag-nssfs-86};
	the second is due to Pach et al.~\cite{pss-ubnpk-92}.
	See~\cite{e-acg-87} and~\cite{aas-olals-97} for history and 
	further references.}
But recently, building on the work of
Agarwal, Aronov, and Sharir~\cite{aas-olals-97}, 
Tamal Dey improved the
upper bound to $O(n k^{1/3})$~\cite{Dey}.  Here I will sketch his 
proof\footnote{
	I will follow his first proof, which employs duality.
	He has since developed a proof that never leaves the primal setting.
}
on an example.

Consider the set of $n=10$ points shown in Fig.~\ref{dey.points}.\footnote{
	I will assume throughout that no three points of $P$ lie on a line,
	and no two points lie on a vertical.  Neither assumption
	affects the worst-case bounds.}
The line shown determines a $5$-set, as there are exactly $5$ points
in the halfplane below.  In this instance, there are also $5$ points above,
so this line determines two $5$-sets.  We will only count the $k$-sets
below their halfplane bounding line, which clearly suffices for an
asymptotic bound; this will be our first reduction.

\begin{figure}[htbp]
\begin{center}
\ \psfig{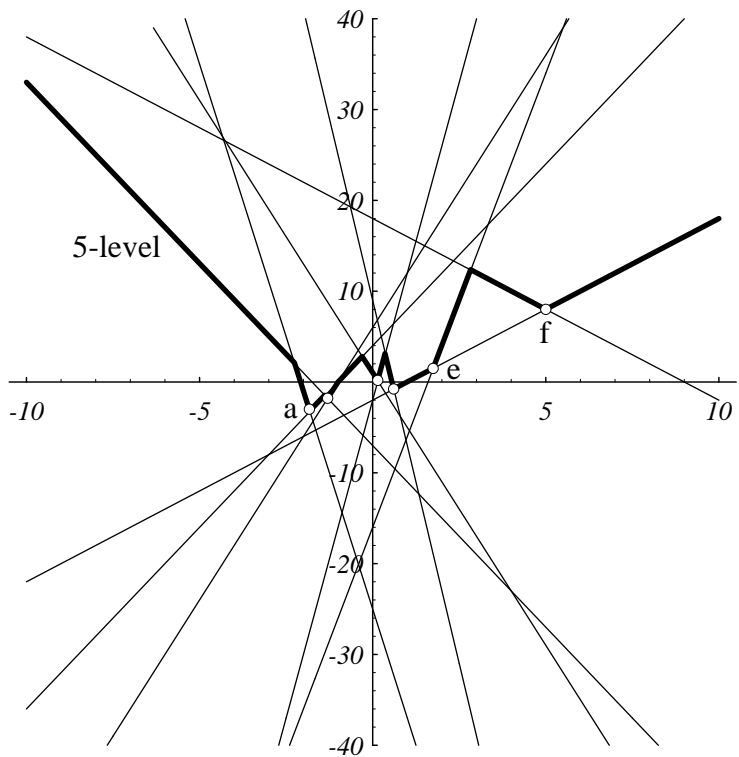}
\end{center}
\caption{The arrangement dual to the points in
Fig.~\protect\ref{dey.points}, with the $5$-level highlighted.
The six vertices of $V_4$ ($a ,\ldots,f$, left to right) are circled.
The other vertices of the $5$-level belong to $V_5$.
}
\label{dey.lines}
\end{figure}

The next step is familiar to those working in combinatorial geometry:
dualize each point of $P$ to a line, transforming the point set to
an arrangement of lines $\A$.  I will use the transformation
$(a,b) \mapsto y=ax-b$ from primal to dual
(and $(a',b') \mapsto y=a'x-b'$ from dual to primal),
which preserves the
above-below relation:  
a point of $P$ is below a line $L$
iff the dual line is below the point dual to $L$.
Define the $k$-level of $\A$ to be
the closure of the set of points of the lines of $\A$ 
that have exactly $k$ lines below them.
The number of vertices on the $k$-level and the number of
$k$-sets (with $k$ points below) differ by at most $1$.
Take any $k$-set line, translate it upward until it hits a point
of $P$, and then rotate it counterclockwise about that point until
it hits another.  This line corresponds to a vertex of the $k$-level.
Fig.~\ref{dey.lines} shows the $5$-level of the arrangement dual
to the points in Fig.~\ref{dey.points}.
Note that not every point of the $5$-level has five lines below:
for example, the rightmost vertex of the level, at $f=(5,8)$, has four
lines below.\footnote{
	This is why the level was defined via closure, to fill
	in these holes.}
Let $V_k$ be the set of vertices of $\A$ with exactly
$k$ lines below.  Then the vertices of the $k$-level are
precisely $V_k \cup V_{k-1}$.  Thus the vertex at $f$ is a member
of $V_4$; all six vertices in $V_4$ are marked with open circles
in the figure.

The second reduction is that it suffices to bound $|V_{k-1}|$, as these
vertices correspond to lines through two points of $P$ that have
exactly $k-1$ points below them.  Perturbing the line corresponding to
a vertex of $V_{k-1}$ results in two $k$-sets below.
Thus the point $f=(5,8)$ in Fig.~\ref{dey.lines} maps to the
line $y=5x-8$ through points $(-1,-18)$ and $(2,2)$ of $P$; the
perturbation that includes the former point but excludes the
latter is the line shown in Fig.~\ref{dey.points}.

\begin{figure}[htbp]
\begin{center}
\ \psfig{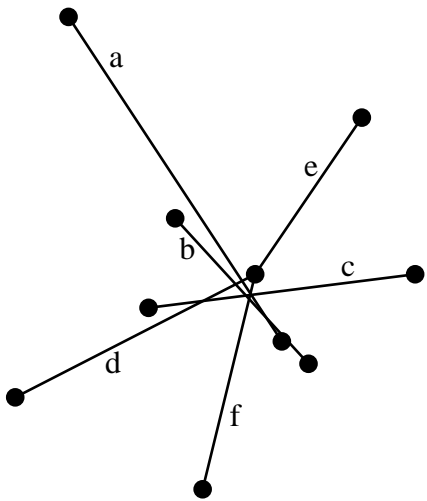}
\end{center}
\caption{The graph $G$ has an edge for every vertex of $V_{k-1}$.
The edge labels match those in Fig.~\protect\ref{dey.lines}.
}
\label{dey.G}
\end{figure}

Now Dey defines a straight-line geometric graph $G$ 
in the primal setting that connects a
pair of points of $P$ with an edge iff the line containing that edge
dualizes to a vertex in $V_{k-1}$.  See Fig.~\ref{dey.G}.  We now seek
to bound the number of edges $t=|V_{k-1}|$ of $G$.
Dey obtained his bound by relating $t$ to the number of crossings $X$
of edges of $G$.
In particular, Ajtai et al.~\cite{aks-scps-83} and Leighton~\cite{l-nlbtv-84}
proved that there must 
be at least $c t^3 / n^2$
crossings among $t > 4n$ edges of a straight-line geometric graph, 
$c$ a constant.
As the case $t \le 4n$ is easy,\footnote{
	Our running example only has $t=6 < 4n=40$.}
this establishes that $c t^3 / n^2 < X$.

\begin{figure}[htbp]
\begin{center}
\ \psfig{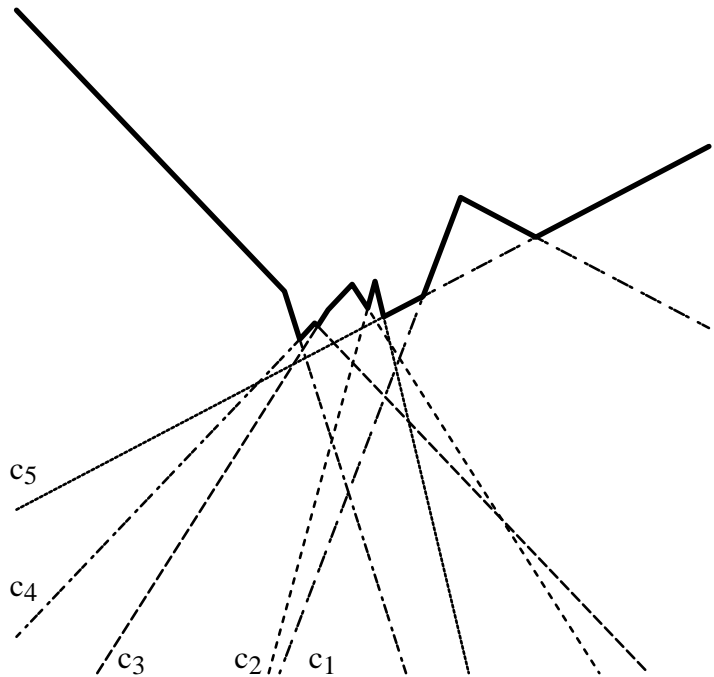}
\end{center}
\caption{The $k=5$ concave chains cover the arrangement below the $k$-level.
}
\label{dey.concave}
\end{figure}

\begin{figure}[htbp]
\begin{center}
\ \psfig{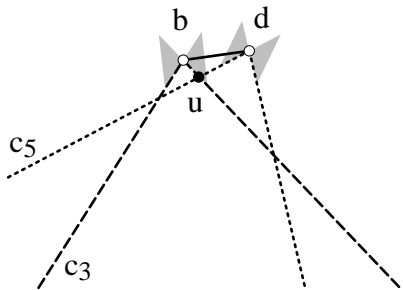}
\end{center}
\caption{The common tangent to chains $c_3$ at vertex $b$
and $c_5$ at vertex $d$ 
(corresponding to
the crossing of edges $b$ and $d$ in $G$ (Fig.~\protect\ref{dey.G}))
is charged to the point $u$
where these two chains cross below the
tangent. Since the common tangents of a pair of $x$-monotone concave chains
have disjoint $x$-spans, this charging is unique.
}
\label{dey.wedge}
\end{figure}

An upper bound for $X$ is obtained by a close analysis of the lines below
the $k$-level, which I will only sketch.  
Agarwal et al. emphasized in~\cite{aas-olals-97} a way of
viewing those lines that is a key part of Dey's argument.  They partitioned
the portion of $\A$ below the $k$-level into a union of $k$ concave
chains $c_1, c_2, \cdots$, which turn only at the vertices of $V_{k-1}$.
Fig.~\ref{dey.concave} will serve in place of a formal definition.
A consequence of our duality is that
a pair of crossing edges in $G$ corresponds to a line tangent
to two concave chains, passing through a vertex of each; 
see Fig.~\ref{dey.wedge}.
This bounds $X$ by the number of such common tangents.
The number of these tangents is in turn bounded from above by the
number of times the chains cross each other,
as explained in the caption to Fig.~\ref{dey.wedge}.
As all of these crossings occur below the $k$-level of $\A$,
there cannot be more than the number of vertices below this level
which, as mentioned previously, is known to be at most $nk$.

So we now have $c t^3 / n^2 < X <  n k$.  Solving for $t$
yields $t = O(n k^{1/3})$.

\vspace{4mm}
\footnotesize
\noindent
{\bf Acknowledgements.}
I thank Pankaj Agarwal, Boris Aronov, Tamal Dey, and Micha Sharir for comments.

\normalsize

\begin{flushright}
(Written 27 July 1997)
\end{flushright}
\end{document}